\documentclass[a4paper,11pt]{article}
\pdfoutput=1 

\usepackage{jcappub} 

\usepackage{aas_macros}
\usepackage{graphicx, soul}
\usepackage{amsmath}
\usepackage{amssymb}

\usepackage[normalem]{ulem}
\usepackage[T1]{fontenc}

\newcommand{\Msun}{\mathrm{M}_{\odot}}
\newcommand{\hcMpc}{h^{-1} \mathrm{cMpc}}
\newcommand{\de}{{\text{d}}}

\newcommand{\fig}[1]{Figure~\ref{#1}}

\title{Efficient hybrid technique for generating sub-grid haloes in reionization simulations}
\author[a,1]{Ankur Barsode}\note{Corresponding author}%
\author[b]{and Tirthankar Roy Choudhury.}

\affiliation[a]{International Centre for Theoretical Science, Tata Institute of Fundamental Research, Bangalore 560089, India}
\affiliation[b]{National Centre for Radio Astrophysics, Tata Institute of Fundamental Research, Pune 411007, India}

\emailAdd{ankur.barsode@icts.tifr.res.in}
\emailAdd{tirth@ncra.tifr.res.in}

\abstract{

    Simulating the distribution of cosmological neutral hydrogen (HI) during the epoch of reionization requires a high dynamic range and is hence computationally expensive. The size of the simulation is dictated by the largest scales one aims to probe, while the resolution is determined by the smallest dark matter haloes capable of hosting the first stars. We present a hybrid approach where the density and tidal fields of a large-volume, low-resolution simulation are combined with small haloes from a small-volume, high-resolution box. By merging these two boxes of relatively lower dynamic range, we achieve an effective high-dynamic range simulation using only 13\% of the computational resources required for a full high-dynamic range simulation. Our method accurately reproduces the one- and two-point statistics of the halo field, its cross-correlation with the dark matter density field, and the two-point statistics of the HI field computed using a semi-numerical code, all within 10\% accuracy at large scales and across different redshifts. Our technique, combined with semi-numerical models of reionization, provides a resource-efficient tool for modeling the HI distribution at high redshifts.

}

\begin{document}

    \date{}

    \maketitle

    \flushbottom

    \section{Introduction}

    Numerical simulations of reionization, particularly those designed for inference with upcoming 21~cm surveys, require a high dynamic range (for recent reviews, see \cite{2022LRCA....8....3G,2022GReGr..54..102C}). At one end, simulation volumes need to be  $\gtrsim 200$~cMpc to model the largest scales probed by these surveys \cite{2014MNRAS.439..725I,2020MNRAS.495.2354K,2023A&A...669A...6G,2024MNRAS.529.3793A}. At the other end, it is necessary to resolve relatively small mass haloes, given that the first galaxies formed in atomically cooled haloes with masses $\sim 10^8 \, \Msun$. Typically, these simulations would require $\gtrsim 10^{11}$ matter particles which, in turn, would demand $\gtrsim 10$~TB of memory. The requirements become even more demanding if one aims to resolve haloes where molecular cooling is efficient. A high number of particles is also necessary for resolving small spatial scales, which is crucial for accurately modeling the sinks of ionizing photons.

    In semi-numerical simulations of reionization, where the radiative transfer physics is replaced by an appropriate photon counting algorithm, a common compromise is to use large-volume simulation boxes with relatively coarse resolution \cite{2007ApJ...669..663M,2011MNRAS.411..955M,2010MNRAS.406.2421S,2014MNRAS.443.2843M,2016MNRAS.457.1550H,2017MNRAS.464.2992M,2018MNRAS.477.1549H,2018MNRAS.481.3821C}. In some extensions to these models, the effect of recombinations is accounted for through sub-grid models \cite{2009MNRAS.394..960C,2014MNRAS.440.1662S}. The coarse resolution of these simulations also implies that they are not able to resolve the small mass haloes. Instead, they use the gridded density fields to compute the conditional halo mass functions within each grid cell to populate them with sub-grid haloes. This approach is typically sufficient because the primary interest lies in modeling the large-scale 21~cm signal pertinent to upcoming experiments, rather than in the detailed properties of the haloes (e.g., their density profiles). Therefore, only the location and mass of the haloes are required for these simulations.

    Such sub-grid prescriptions for populating small mass haloes in coarse resolution simulations have been widely used for modeling the epoch of reionization. One of the early methods was based on using conditional Press-Schechter mass function \cite{1991ApJ...379..440B} scaled to match the Sheth-Tormen mass function \cite{1999MNRAS.308..119S}, which was found to provide a good match to the halo bias from $N$-body simulations \cite{2004ApJ...609..474B}. This method has been widely used in semi-numerical models of reionization for populating haloes that are otherwise unresolved in the box \cite{2007ApJ...654...12Z,2008MNRAS.384.1069B,2011MNRAS.411..955M,2011A&A...527A..93S,2015MNRAS.447.1806G}. A slightly alternate approach has been to use the analytical conditional mass function based on the ellipsoidal collapse to populate the unresolved haloes \cite{2018MNRAS.481.3821C,2022A&A...667A.118D}. Alternatively, a fit to the conditional mass function, obtained from a high-resolution simulation capable of resolving the smallest haloes forming stars \cite{2012ApJ...756L..16A,2015MNRAS.450.1486A} has also been used. Obvious extensions to these techniques involve introducing scatter in the halo numbers \cite{2011MNRAS.411..955M,2022A&A...667A.118D}.

    The conditional halo mass functions based on analytical calculations or functional fits encounter several challenges. For instance, the analytical relations require knowledge of the Lagrangian density as opposed to the Eulerian ones obtained from the simulations. Usually, the Eulerian to Lagrangian mapping is done using spherical approximation \cite{2002MNRAS.329...61S}, which may not be accurate when the grid cell sizes become small. To bypass this reliance on analytical relations, an alternate approach could be to combine information from two independent simulation boxes, one high-resolution but small volume, and another coarse resolution with large volume. The motivation behind such hybrid approaches comes from the fact that computational cost increases with the dynamic range, so two simulations with narrow dynamic ranges can be quite inexpensive as compared to simulating with the full dynamic range. Such methods have been used to model the density and velocity fields during the late stages of reionization \cite{2015MNRAS.446..566M,2015MNRAS.452..261C}, where they combined a large-scale dark-matter-only simulation with a high-resolution hydrodynamical simulation to obtain a large-scale, high-resolution HI density field.

    In this work, we refine the hybrid method based on multiple simulation volumes to populate low-mass haloes in the coarse resolution box that are otherwise unresolved. Our method does not require any analytical fits to the conditional mass functions or any mapping to the Lagrangian field, rather depending on matching similar grid cells from the two boxes. We also improve on the existing methods in the sense that we use not only the density to condition the mass function but also the tidal field. As it turns out, using the tidal information leads to more robust conditional mass functions in agreement with the actual one. We restrict our analysis to high redshifts $z \gtrsim 6$ and evaluate the success of our method with respect to how accurately it can produce the large-scale ionized hydrogen field.

    The outline of the paper is as follows: The following section illustrates our methodology for constructing the hybrid box containing the sub-grid haloes. We demonstrate the accuracy of our method by comparing it with a full high-dynamic range simulation in section \S \ref{sec:results_discussion}. Finally, we conclude in \S \ref{sec:summary} with a brief discussion of the advancements our method can bring to the studies of reionization.

    \section{Methodology}
    \label{sec:methodology}

    \begin{figure*}
        \centering
        \includegraphics[width=0.95\textwidth]{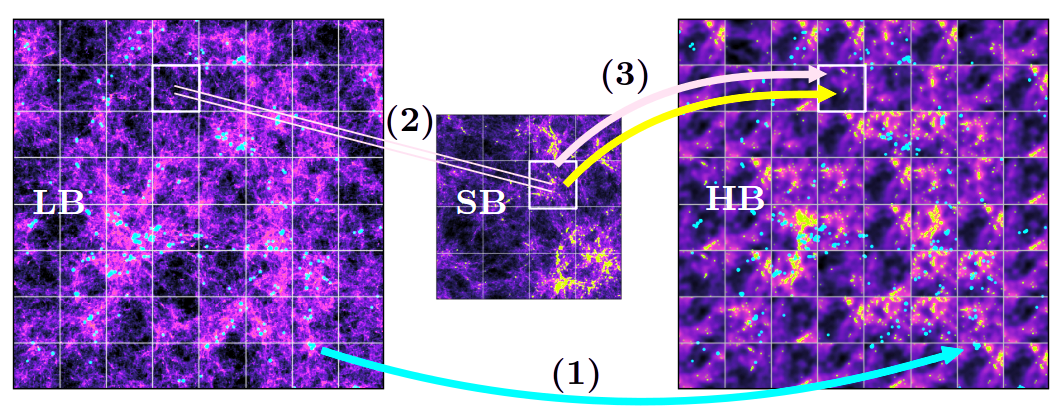}
        \caption{
            Illustration of the methodology to construct the hybrid box: (1) high-mass haloes (\textit{cyan}) from the large low-resolution box (LB) are directly copied to the hybrid box (HB), (2) for each grid cell in LB, best matching tiles are found in a small high-resolution box (SB) using either the density field or the eigenvalues of the tidal tensor field, (3) the low-mass haloes (\textit{yellow}) and dark matter density field smoothed at the grid scale (\textit{lavender}) are picked up from these locations and substituted in the hybrid box, ensuring that the large scale density field in HB is identical to LB.
        }
        \label{fig:DMH_methodology}
    \end{figure*}

    In this section, we describe our algorithm for constructing the hybrid simulation boxes in detail.

    Let us assume that we want to construct a hybrid box equivalent to a (cubical) simulation box of volume $V_\mathrm{box} = L_\mathrm{box}^3$ with $N_\mathrm{part}$ particles. The smallest halo that can be resolved in this simulation would depend on the particle mass resolution $M_\mathrm{part} = (V_\mathrm{box} / N_\mathrm{part})~\bar{\rho}_m$, where $\bar{\rho}_m$ is the mean density of matter. Our procedure for constructing the hybrid box requires two simulations of relatively coarser resolution, one having a size as large as the target simulation and another having a mass resolution the same as the target one. The details are as described in the following:

    \begin{itemize}

        \item \textbf{Large Box (LB):} This simulation box has a volume $V_\mathrm{box, large} = V_\mathrm{box}$ corresponding to the ultimate volume of the simulation we want to construct. Let $N_\mathrm{part, large}$ be the number of particles in this simulation. Clearly, if we want this simulation to be less expensive than the target simulation, it must contain fewer particles than the target one, i.e., $N_\mathrm{part, large} < N_\mathrm{part}$.

        \item \textbf{Small Box (SB)}: This simulation box has a volume $V_\mathrm{box, small}$ with $N_\mathrm{part, small}$ number of particles, chosen such that the mass resolution is equal to the target simulation, i.e.,
        \begin{equation}
            \dfrac{V_\mathrm{box, small}}{N_\mathrm{part, small}} = \dfrac{V_\mathrm{box}}{N_\mathrm{part}}.
        \end{equation}
        As with the LB, we want $N_\mathrm{part, small} < N_\mathrm{part}$, hence $V_\mathrm{box, small} < V_\mathrm{box}$. This also implies that the mass resolution for SB is finer than the LB, i.e., $M_\mathrm{part, small} < M_\mathrm{part, large}$.
    \end{itemize}

    The simulation outputs are used to compute the cloud-in-cell (CIC)-smoothed density contrast $\delta_m(\mathbf{x}) \equiv \rho_m(\mathbf{x}) / \bar{\rho}_m - 1$ on a uniform grid. The grid size $\Delta x_\mathrm{CIC}$ for each box is chosen to be equal to the mean inter-particle distance in that box.

    However, to make the semi-numerical algorithms more efficient, it is often the case that the ionization fields are generated at a relatively coarser resolution. Denoting this target grid size as $\Delta x_\mathrm{grid}$ ($\geq \Delta x_\mathrm{CIC}$ for either box), we smooth the high-resolution density field to this resolution using the boxcar filter. The number of grid cells for the two simulation boxes is given by
    \begin{equation}
        N_\mathrm{grid, small} = \frac{V_\mathrm{box, small}}{\Delta x_\mathrm{grid}^3} < \frac{V_\mathrm{box, large}}{\Delta x_\mathrm{grid}^3} = N_\mathrm{grid, large}.
    \end{equation}

    This smoothed density field is used for computing the tidal tensor field
    \begin{equation}
        \nabla^2 T_{ij}(\mathbf{x}) = \frac{\partial^2 \delta_m(\mathbf{x})}{\partial x_i \partial x_j},~~~ i,j = 1,2,3.
    \end{equation}
    In practice, we compute the derivative with respect to the spatial coordinates $x_i$ in Fourier space and then transform it back to the configuration space. The main quantities of our interest are the three eigenvalues $\lambda_1(\mathbf{x}), \lambda_2(\mathbf{x}), \lambda_3(\mathbf{x})$ of $T_{ij}(\mathbf{x})$, ordered such that $\lambda_1(\mathbf{x}) \leq \lambda_2(\mathbf{x}) \leq \lambda_3(\mathbf{x})$. Note that the density contrast $\delta_m(\mathbf{x}) = \lambda_1(\mathbf{x}) + \lambda_2(\mathbf{x}) + \lambda_3(\mathbf{x})$ is the sum of these eigenvalues of the tidal tensor.

    We further identify and generate the halo catalogue for each of the two boxes, using the Friend-of-Friends (FoF) algorithm \cite{1985ApJ...292..371D}. The smallest halo $M_{h, \mathrm{min}}$ used in this work is assumed to have 20 particles. Note that this minimum halo mass would be smaller in SB than in LB $M_{h, \mathrm{min, small}} < M_{h, \mathrm{min, large}}$, since SB has a resolution finer than LB. Also, since the resolution for SB is chosen to be identical to that of the target box, we have $M_{h, \mathrm{min, small}} = M_{h, \mathrm{min}}$. Once identified, we treat the haloes as point masses having a mass equal to that of the halo, and located at the center of mass of its constituent particles.

    To compute the power spectrum of the halo field, its cross-spectrum with the density field, and ultimately the ionization field (to be discussed later), we also need to transform these point mass haloes into a gridded halo field $\delta_h(\mathbf{x})$. This is done by first CIC-depositing the haloes for the two boxes at the fine resolution of $\Delta x_\mathrm{CIC}$, and then boxcar-smoothing them to the target resolution $\Delta x_\mathrm{grid}$.

    We also run another simulation, called \textbf{Reference Box (RB)}, which is used for comparing the results obtained from our hybrid algorithm (discussed in the next section \S \ref{sec:results_discussion}). The specifications of this simulation are identical to the ones we wish to achieve using the hybrid algorithm. Specifically, it has a volume $V_\mathrm{box} = L_\mathrm{box}^3$ with $N_\mathrm{part}$ particles. The particle positions for this box are CIC-smoothed to obtain the density field, which is then boxcar-smoothed to a coarse resolution $\Delta x_\mathrm{grid}$ for later use. The haloes are identified using the same FoF algorithm as above and the halo field at the desired resolution is computed in a manner identical to those for the LB and SB. Note that this box does \emph{not} play any role in the implementation of the hybrid algorithm, it is only used for validating the results obtained from the hybrid algorithm. We also control the seed for the initial realization of the cosmological fields for the RB so that the large-scale fields are identical to the LB. This ensures that the effects of cosmic variance are minimized at scales comparable to $L_\mathrm{box}$.

    In this work, we choose to work with the case where,
    \begin{equation}
        N_\mathrm{part, small} = N_\mathrm{part, large} = \frac{N_\mathrm{part}}{8},
    \end{equation}
    which implies that both LB and SB contain 8 times fewer particles than the target simulation. This also implies the following relations between the box lengths
    \begin{equation}
        L_\mathrm{box, small} = \frac{L_\mathrm{box, large}}{2} = \frac{L_\mathrm{box}}{2},
    \end{equation}
    the number of grid cells along each direction
    \begin{equation}
        N_\mathrm{grid, small} = \frac{N_\mathrm{grid, large}}{8},
    \end{equation}
    and the minimum halo mass
    \begin{equation}
        M_{h, \mathrm{min, small}} = \frac{M_{h, \mathrm{min, large}}}{8} = M_{h, \mathrm{min}}.
    \end{equation}

    For the analyses in this paper, dark-matter-only simulations were run using \texttt{GADGET-2}\footnote{\url{https://wwwmpa.mpa-garching.mpg.de/gadget/}} \cite{2005MNRAS.364.1105S}. A flat $\Lambda$CDM cosmology was assumed with $h = 0.678, \Omega_m = 0.308, \Omega_b = 0.0, \sigma_8 = 0.829, n_s = 0.961$. The initial conditions were generated at $z=60$ using \texttt{N-GenIC} \cite{2015ascl.soft02003S}.
    The initial power spectrum was generated using the analytical fit of Eisenstein \& Hu \cite{1999ApJ...511....5E}. The details of the simulations used to illustrate the details of our method are mentioned in Table~\ref{tab:simulations}. In addition to these, we also test our method using other box sizes in \S\ref{sec:L_40_160}. Our RB contains $1024^3$ particles, the maximum we can use given the memory available to us. We choose the box size $80 \hcMpc$ which is slightly smaller than what is ideal for modeling the 21~cm signal, and it allows us to probe halo masses an order of magnitude larger than the atomically cooled haloes. The SB and LB each contain $512^3$ particles, hence requiring almost an order of magnitude less memory than RB. The total combined run time for SB and LB is $\sim 7$ times less than RB.

    \begin{table}
        \begin{tabular}{lccccc}
            \hline
            \hline
            Name & $L_\mathrm{box}$ & $N_\mathrm{part}$ & $M_{h, \mathrm{min}}$ & CPU time$^{a}$ & Max RAM\\
            & ($\hcMpc$) & & ($h^{-1} \Msun$) & (hours) & (GB)\\
            \hline
            \textbf{Large Box (LB)} & 80 & $512^3$ & $6.52 \times 10^9$ & $\approx 2.1 \times 10^2$ & $\approx 20$\\
            \textbf{Small Box (SB)} & 40 & $512^3$ & $8.15 \times 10^8$ & $\approx 2.2 \times 10^2$ & $\approx 20$\\
            \hline
            \textbf{Reference Box (RB)} & 80 & $1024^3$ & $8.15 \times 10^8$ & $\approx 2.9 \times 10^3$ & $\approx 160$\\
            \hline
            \hline

        \end{tabular}

        $^a$simulations were run only up to $z = 5$.

        \caption{Specifications of the $N$-body simulations used to construct the hybrid boxes.}
        \label{tab:simulations}

    \end{table}

    \subsection{Constructing the hybrid box}

    The LB contains the large-scale density field on a uniform grid of size $\Delta x_\mathrm{grid}$, which is to be used for generating the ionization field. However, this simulation can resolve haloes only above a mass $M_{h, \mathrm{min, large}} = 6.52 \times 10^9 h^{-1} \Msun$, while our aim is to resolve haloes as small as $M_{h, \mathrm{min}} = 8.15 \times 10^8 h^{-1} \Msun$, as obtained in the RB. It is possible to resolve these haloes in the SB which, however, is unable to access the largest scales we are interested in. To obtain the conditional halo mass function for $M_h \geq M_{h, \mathrm{min}}$ at every grid cell in LB, we use the grid cells from the SB. This implicitly assumes that the conditional mass function is determined primarily by local conditions. Under such an assumption, for a given grid cell in LB, one only needs to find cells in the SB that have similar properties.

    Now, regarding which properties could be appropriate, we consider two options:
    \begin{enumerate}
        \item \textbf{Density:} The first option is that we find a grid cell in the SB which have a matter density contrast $\delta_m$ similar to the cell in LB under consideration. Once such a match is identified, we simply populate all the haloes with masses between $M_\mathrm{min, small}$ and $M_\mathrm{min, large}$ in the cell in LB. This avoids any double-counting, as all haloes with $M_h \geq M_\mathrm{min, large}$ are already present in the LB. It is likely that the match between the $\delta_m$'s from the LB and SB would not be perfect, so to conserve the mass within haloes, we scale the masses of all these sub-grid haloes by the ratio of the densities. This leads to a hybrid box with a population of haloes with $M_h \geq M_{h, \mathrm{min, small}}$.

        \item \textbf{Tidal tensor:} We also explore another possibility that the conditional halo properties may not be determined solely by the local density, but also by the tidal environment. To account for this possibility, we use the eigenvalues $\lambda_i$ of the tidal tensor to find the matching grid cells between LB and SB. While trying to identify the cell with the best match, we simply find one having the nearest value of $\sum_{i=1}^3 (\lambda_i^\mathrm{LB} - \lambda_i^\mathrm{SB})^2$. Following this identification of the best-matched cell, we populate the sub-grid haloes in a manner identical to the density-based method described above. In particular, in this case, too, we scale the halo masses by the ratio of the cell densities.
    \end{enumerate}
    A schematic representation of the above algorithm can be found in \fig{fig:DMH_methodology}.

    The above method provides sub-grid haloes in the density field of the LB, which forms our \textbf{Hybrid Box (HB)}. Note that the above methods would lead to discontinuities in the halo field at the boundaries of the grid cells, hence the correlation functions would show a lack of correlation at scales $k \gtrsim 2 \pi / \Delta x_\mathrm{grid}$. However, both the above methods are expected to preserve the density-halo correlation at larger scales $k \lesssim 2 \pi / \Delta x_\mathrm{grid}$. We will discuss these in more detail in the next section \S \ref{sec:results_discussion} by comparing the results from the HB with the RB.

    \subsection{Ionization field}
    \label{sec:script}
    Our main quantity of interest is the ionization field. We use the semi-numerical model \texttt{SCRIPT}\footnote{\url{https://bitbucket.org/rctirthankar/script}} \cite{2018MNRAS.481.3821C, 2022ascl.soft04013C} to generate the ionization fields for both HB and RB. The algorithm requires a density field in a rectangular grid, along with haloes that are capable of producing ionizing photons. We assume the ionizing photons produced in a halo to be proportional to the halo mass $M_h$, with the constant of proportionality being denoted by the ionization efficiency $\zeta$. We also assume that all haloes with $M_h \geq M_{h, \mathrm{min}}$ are capable of forming stars and contributing to the ionizing photon budget.

    The algorithm of \texttt{SCRIPT} conserves the number of photons explicitly and hence provides large-scale ionization maps that are independent of the grid size $\Delta x_\mathrm{grid}$ used for generating the ionization fields. Given the density field on a grid and the number of ionizing photons produced by haloes in a grid cell, the algorithm assigns ionized regions of appropriate volumes around the ‘source’ cells. This process is repeated independently for all cells containing sources in the box, and it may result in some grid cells that receive photons from multiple source cells being ‘overionized’. These excess ionizing photons in these unphysical overionized cells are subsequently distributed among the surrounding neighboring cells which are yet to be fully ionized. The process is continued until all the overionized cells are properly accounted for.

    \section{Results and Discussion}
    \label{sec:results_discussion}

    In this section, we compare our hybrid method for generating sub-grid haloes with the reference box. We compare our results for three choices of the grid size, namely, $\Delta x_\mathrm{grid} = 1.25$, $0.625$, and $0.3125~\hcMpc$, which correspond to $N_\mathrm{grid} = 64^3$, $128^3$ and $256^3$, respectively. Note that smaller grid sizes will allow us to probe the power spectra at higher $k$-values. However, such grids would contain relatively fewer haloes, and hence the variance would be larger. Consequently, this may lead to more discrepancies between the distribution of haloes in the cells in HB compared to the RB. We choose the reference redshift as $z = 7$ for most of the results, but we also show the evolutionary trends of our scheme in the latter part.

    \subsection{Matter fields}

\begin{figure}
        \includegraphics[width=\columnwidth]{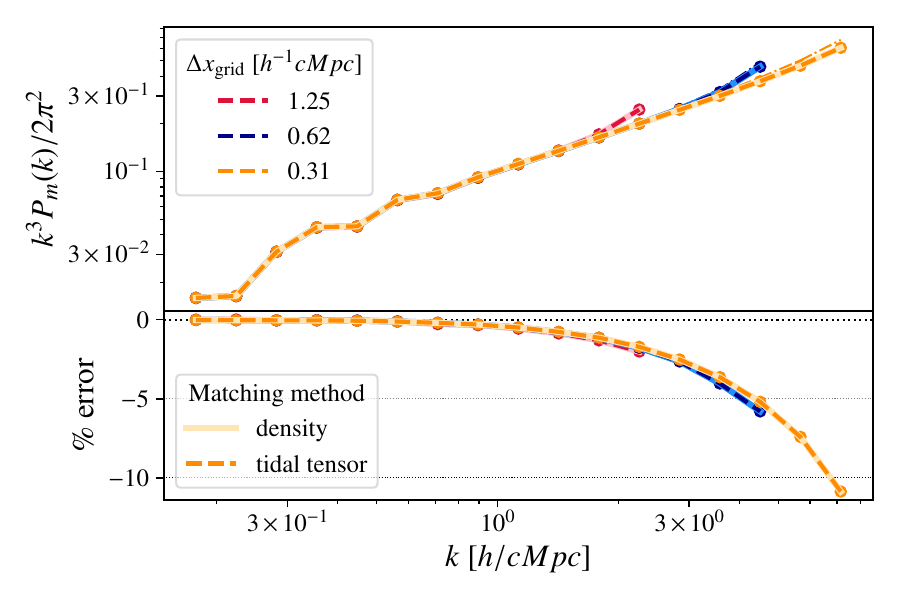}
        \caption{Comparison of hybrid dark matter density power spectra at $z=7$ with the reference simulation (\textit{thin dash-dot lines}) for different grid sizes and matching methods as mentioned in the legends. The lower panel shows the \% error in each power spectrum with respect to RB. The power spectra mostly overlap with RB, though they show $\sim~10\%$ deviation at large k. However, since LB and HB are designed to have identical spectra, this deviation must be entirely due to differences in LB and RB arising from different force resolutions. We show that this is indeed the case by plotting the power spectrum of LB using \textit{filled circles with the same color as the corresponding HB}.}
        \label{fig:hb_Ngrid_512_z_7.0_mm}
    \end{figure}

    Let us first compare the gridded density fields of HB and RB. \emph{Note that the density field for the HB is simply that of LB}, and the details of the hybridization scheme do not play any role here. The power spectra $P_m(k)$ (in units of $h^3\mathrm{cMpc}^{-3}$) of the density contrasts $\delta_m$ for the two boxes, and show their non-dimensionalized versions $k^3P_m(k)/2\pi^2$ in \fig{fig:hb_Ngrid_512_z_7.0_mm}. Their relative discrepancy is shown in the lower panel of the same figure. The lowest $k$ we consider for the plot is $\sim 2 \pi / (40 \hcMpc)$ as smaller $k$-modes show effects of aliasing arising from the box length of $80 \hcMpc$. The maximum $k$ used depends on the grid size $\Delta x_\mathrm{grid}$.

    It is clear that the power spectra are identical at large scales $k \lesssim 1 h/$~cMpc. This is because we have controlled the seeds for generating the initial density field of the RB and LB such that they are identical at large scales. At smaller scales, the power of LB is lower than the RB, a direct consequence of the lower force resolution of LB. At $k \gtrsim 4 h/$~cMpc, the difference exceeds $\sim 5\%$.

    \subsection{Hybrid halo catalogue}

    \begin{figure}
        \includegraphics[width=\columnwidth]{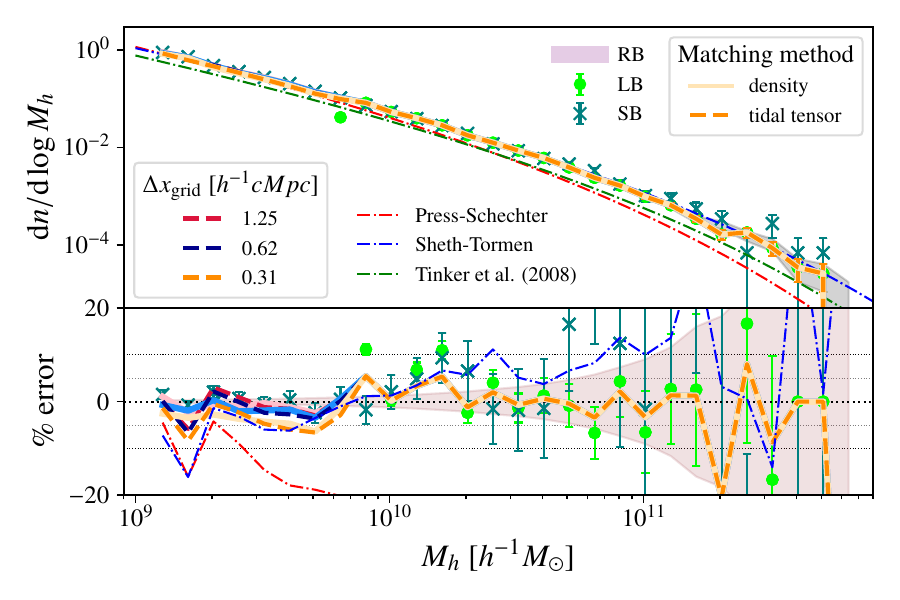}
        \caption{Halo mass functions at $z=7$ of hybrid boxes with different grid sizes and matching methods, along with those of RB, LB, and SB, as indicated by the legends. The error bars on each mass function are calculated assuming the number of haloes in each mass bin follows a Poisson distribution. Semi-analytical mass functions by Press \& Schechter \cite{1991ApJ...379..440B}, Sheth \& Tormen \cite{1999MNRAS.308..119S}, and Tinker et. al. \cite{tinker2008toward} are also shown. The lower panel shows the \% error in the mean estimate of each mass function with respect to that of the RB.}
        \label{fig:hb_Ngrid_512_z_7.0_HMF}
    \end{figure}

    The most important outcome of our method is the high dynamic range hybrid halo catalogue. Figure \ref{fig:hb_Ngrid_512_z_7.0_HMF} shows the mass function of the haloes in the HB, compared against that from RB, at a redshift of 7. For comparison, we also show the analytical mass functions of Press \& Schechter \cite{1991ApJ...379..440B}, Sheth \& Tormen \cite{1999MNRAS.308..119S}, and Tinker et. al. \cite{tinker2008toward}. We find that the mass function of the RB agrees best with the Sheth-Tormen mass function.

    We also show the mass functions obtained from LB and SB. The RB covers a wide range of haloes in the range $10^9 h^{-1} \Msun \lesssim M_h \lesssim 5 \times 10^{11} h^{-1} \Msun$. In contrast, the LB haloes are limited to $M_h \gtrsim 10^{10} h^{-1} \Msun$, while the number of high mass haloes ($M_h \gtrsim 10^{11} h^{-1} \Msun$) in the SB become rarer, as is clear from the size of the error-bars. The HB, which is constructed using both LB and SB, provides a mass function that is very similar to the RB. At low masses $M_h \lesssim 10^{11} h^{-1} \Msun$, the agreement is within $10\%$. At large masses, the match is less accurate due to Poisson fluctuations caused by the limited number of haloes. It is important to note that this discrepancy is solely due to the differences in the LB and RB and does not directly affect our hybridization scheme. The agreement between the halo mass functions of HB and RB has little to no dependence on the matching criteria of the cells (density/tidal tensor) or the grid sizes.

    \begin{figure}
        \includegraphics[width=\columnwidth]{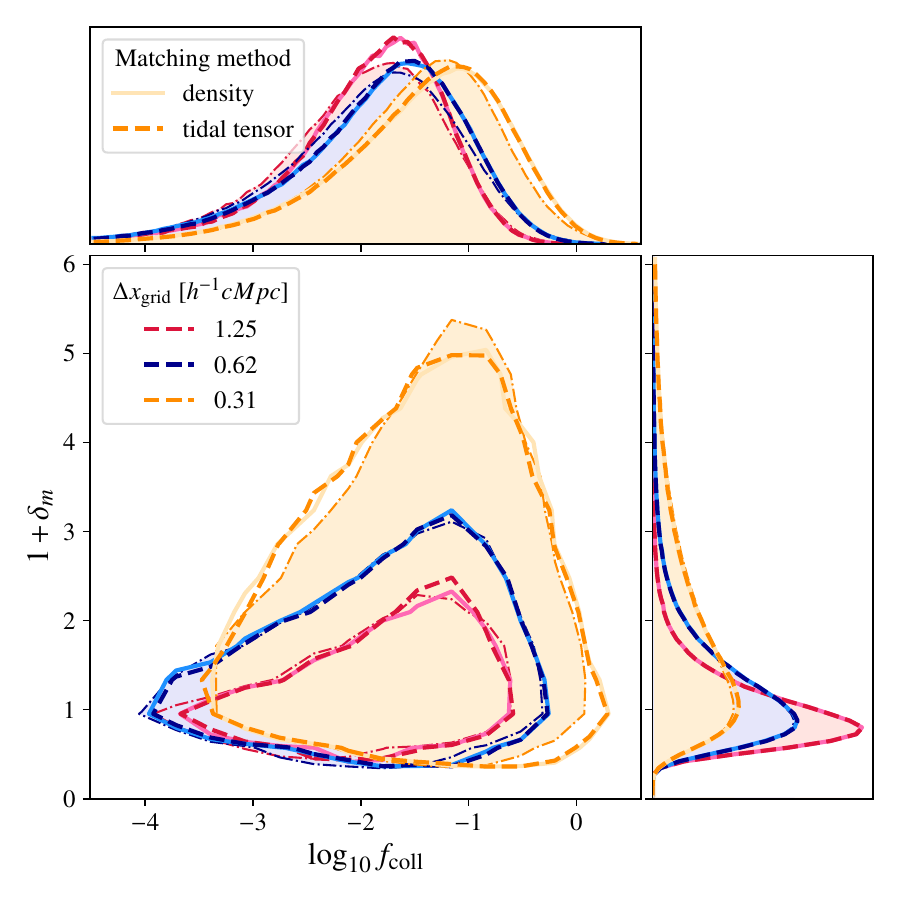}
        \caption{Joint probability distribution of the collapsed fraction $f_\mathrm{coll}$ and density $1 + \delta_m$ for hybrid boxes (\textit{solid/dashed lines} corresponding to density/tidal-tensor matching criteria) as well as reference boxes smoothed at the same resolution as the hybrid box (\textit{light shaded regions surrounded by dot-dash lines}). The central panel shows the 90\% credible region of the joint probability distribution, while the upper and side panels show a marginalized probability distribution of $\log_{10}{f_\mathrm{coll}}$ and $1+\delta_m$ respectively. Places with $\log_{10}{f_\mathrm{coll}} > 0$ are rare artifacts caused by our assumption that haloes are point masses rather than extended objects. This sometimes makes it appear that a grid cell contains more particles than it should on average.}
        \label{fig:delta_vs_fcoll}
    \end{figure}

    Given that the HB produces a good match to the total halo mass function, we next check if the haloes are properly correlated with the underlying density field. We compute the joint distribution of the fraction of mass collapsed to form haloes $f_\mathrm{coll} = \rho_m^{-1} \int_{M_\mathrm{min}}^{\infty} \de M_h~M_h~\de n / \de M_h$, and the density contrast $\delta_m$. The results are shown in \fig{fig:delta_vs_fcoll}. As is clear, the HB captures qualitative features of the correlation between $\delta_m$ and $f_\mathrm{coll}$ as found in RB. Furthermore, the quantitative agreement between the two for grid sizes $\Delta x_\mathrm{grid} \geq 0.625 \hcMpc$ too is excellent. There are some minor disagreements between the HB and RB at low $\delta_m$ and low $f_\mathrm{coll}$, which could probably arise from fewer haloes in low-density cells that are not well captured in the SB. However, these cells contribute negligibly to the ionization photon budget and hence are expected to play a relatively minor role in the ionization field. The disagreement in the joint distribution for $\Delta x_\mathrm{grid} = 0.3125 \hcMpc$ indicates that the method performs less reliably for smaller grid sizes, possibly due to the large variance in halo numbers. We also find that the matching criteria of the grid cells (tidal tensor/density) do not significantly affect the joint distribution.

    \begin{figure}
        \includegraphics[width=\columnwidth]{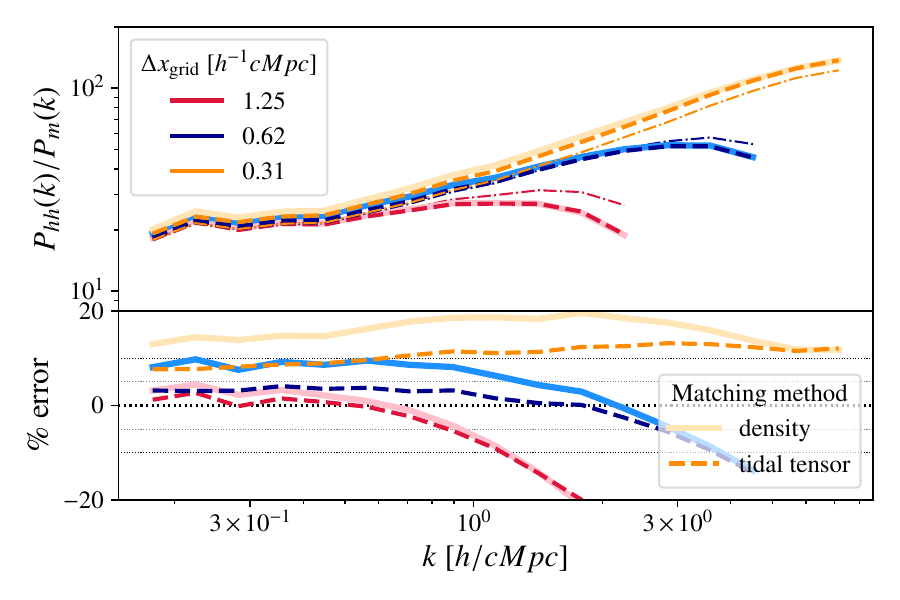}
        \caption{Power spectrum $P_{hh}(k)$ of the halo overdensity field scaled by the density power spectrum $P_m(k)$ at $z=7$ for three values of the grid size $\Delta x_\mathrm{grid}$ and two matching methods as shown in the legend. \textit{Thin dash-dot lines} represent the halo power spectrum for the RB. The lower panel shows the \% error in each scaled power spectrum with respect to the RB.}
        \label{fig:hb_Ngrid_512_z_7.0_hh}
    \end{figure}

    We now turn to the two-point function of the haloes. \fig{fig:hb_Ngrid_512_z_7.0_hh} shows the auto power spectrum $P_{hh}(k)$ of the halo overdensity field scaled by the density power spectrum $P_m(k)$. We find that unlike the halo mass function or the $f_\mathrm{coll} - \delta_m$ joint distribution, the power spectra of HB built using the closeness of the tidal tensor eigenvalues as matching criteria agree better with the RB than those built using the closeness of the density. This indicates the importance of the tidal environment of haloes in shaping their large-scale distribution. We also find that the match between HB and RB is worse for the smallest grid size $\Delta x_\mathrm{grid} = 0.3125 \hcMpc$ we use, disagreements being $\gtrsim 10\%$ at $k \gtrsim 0.5 h/$~cMpc. This is consistent with the results of the $\delta_m - f_\mathrm{coll}$ joint distribution. For the other two grid sizes, the accuracy is within 5\% at large scales, but disagreements increase for small scales. This is because of both the lack of power in the density field for the LB, and also the possible discontinuities in the halo field across the grid cells boundaries arising due to the haloes being substituted from independent grids of the SB.

    \begin{figure}
        \includegraphics[width=\columnwidth]{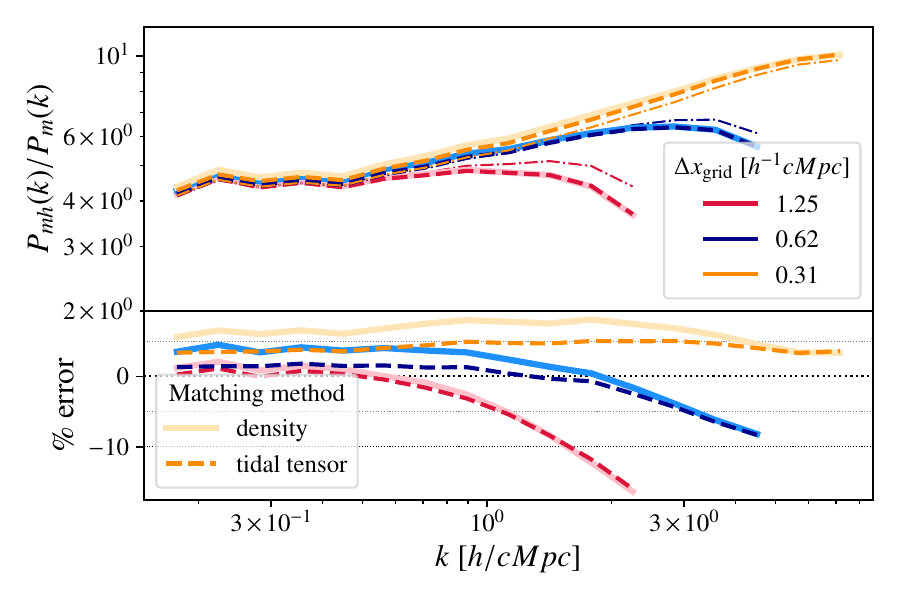}
        \caption{Cross power spectrum $P_{mh}(k)$ of the halo overdensity field with matter density, scaled by the density power spectrum $P_m(k)$ at $z=7$ for three values of the grid size $\Delta x_\mathrm{grid}$ and two matching methods. \textit{Thin dash-dot lines} represent the halo power spectrum for the RB. The lower panel shows the \% error in each scaled power spectrum with respect to the RB.}
        \label{fig:hb_Ngrid_512_z_7.0_mh}
    \end{figure}

    Similar conclusions can be drawn from the cross power spectra $P_{mh}(k)$ of the density and halo fields. We show the ratio $P_{mh}(k) / P_m(k)$ in \fig{fig:hb_Ngrid_512_z_7.0_mh}. The figure shows clear agreement between HB and RB that is within $5\%$ at small $k$ for the larger grid sizes $\Delta x_\mathrm{grid} = 0.625$ and $1.25 \hcMpc$. These results related to the power spectra imply that the HB not only has the right number of haloes, but they are also in the right places with respect to the density field.

    \subsection{Hybrid HI fields}
    \label{sec:HI_fields}
    We next compare the HI density field obtained from the HB with that from the RB. It is worth emphasizing that to get the correct HI field, the halo catalogue needs to be complete down to the lowest mass that can produce ionizing photons, and the haloes must be correctly correlated with the underlying dark matter density field. As indicated by our results so far, we indeed have reasonably accurate density and halo fields to be able to build HI maps. As outlined in \S\ref{sec:script}, we feed the hybrid density and halo fields to \texttt{SCRIPT} to generate the ionization field $1-x_\mathrm{HI}$, and subsequently, the HI field $x_\mathrm{HI} \cdot (1 + \delta_m)$. The HI power spectrum $P_\mathrm{HI}(k)$, in units of $h^3\mathrm{cMpc}^{-3}$, is then computed from the HI density contrast $\delta_\mathrm{HI} = x_\mathrm{HI} \cdot (1 + \delta_m) / \langle x_\mathrm{HI}\rangle_m - 1$, where $\langle x_\mathrm{HI}\rangle_m = \langle x_\mathrm{HI} \cdot (1+\delta_m) \rangle$.

    While comparing the results at $z=7$, we tune the ionization efficiency $\zeta$ in HB such that the mass-averaged global neutral fraction $\langle x_\mathrm{HI}\rangle_m$ is equal to that in RB. This tuning is required because the hybrid boxes do not recover the collapsed field perfectly, which results in the simulation requiring slightly higher or lower ionization efficiency for reaching the same degree of ionization, though the discrepancy is $\lesssim 5\%$. We shall discuss this more towards the end of \S\ref{sec:L_40_160}.

    \begin{figure*}
        \includegraphics[width=\textwidth]{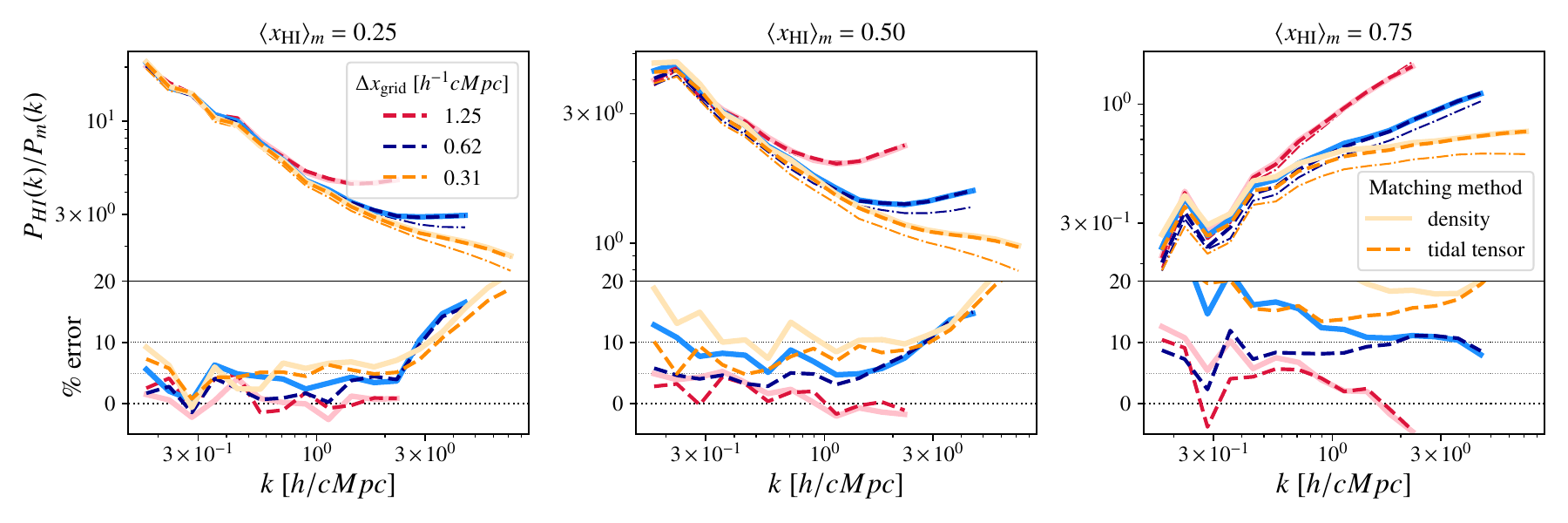}
        \caption{Power spectrum $P_\mathrm{HI}(k)$ of the HI density field scaled by the density power spectrum $P_m(k)$ at $z=7$. Results are shown for three mass-averaged neutral fractions $\langle x_\mathrm{HI}\rangle_m$, obtained by appropriately tuning the ionization efficiency $\zeta$ for three values of the grid size $\Delta x_\mathrm{grid}$ and two matching methods. \textit{Thin dash-dot lines} represent the HI power spectrum for the RB. The lower panel shows the \% error in each scaled power spectrum with respect to the RB.}
        \label{fig:hb_Ngrid_512_z_7.0_HI}
    \end{figure*}

    \begin{figure*}
        \includegraphics[width=\textwidth]{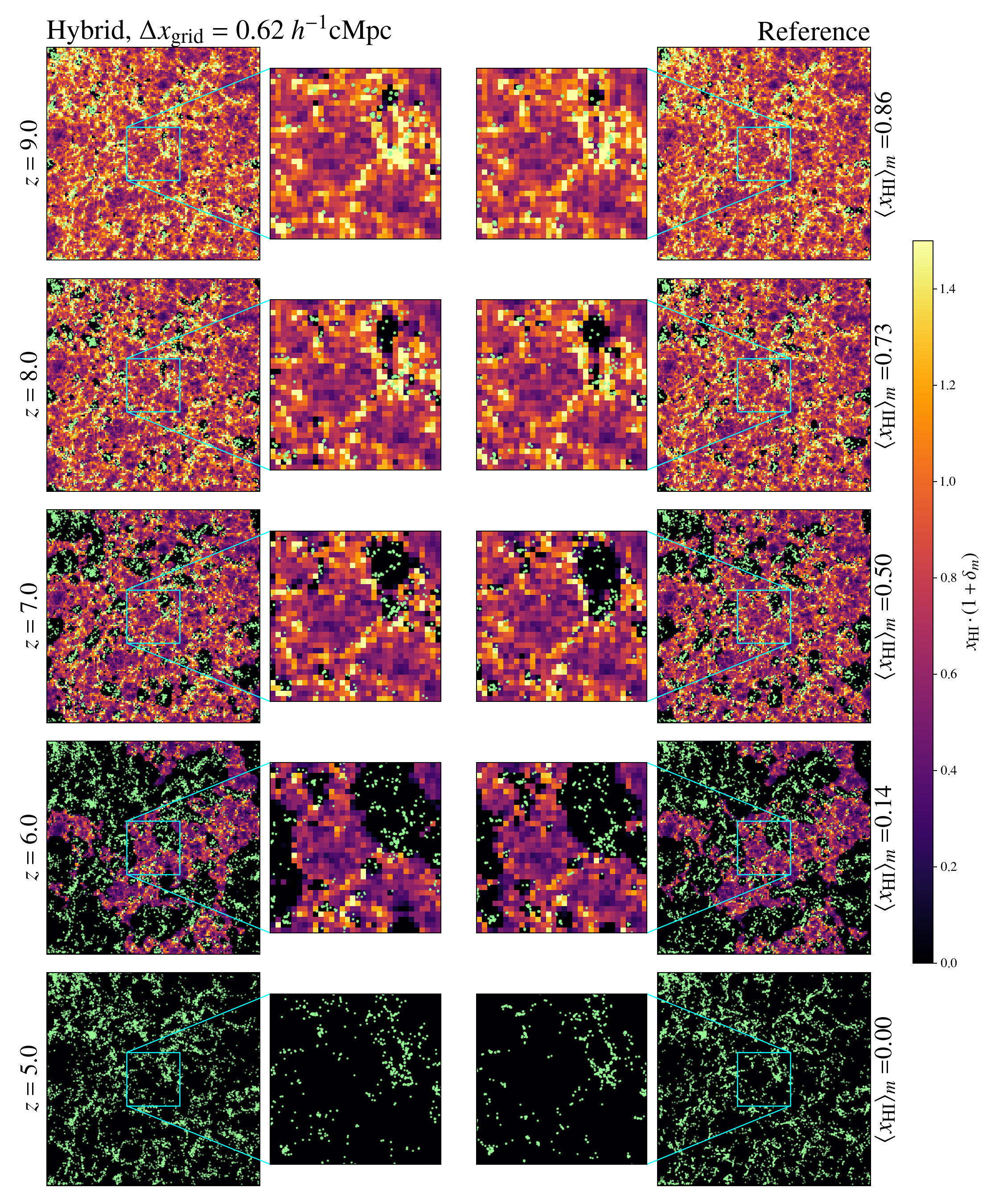}
        \caption{Evolution of HI and halo fields with redshift/mass averaged neutral fraction $\langle x_\mathrm{HI}\rangle_m$. The ionization efficiency $\zeta$ was chosen such that the reference box (RB) had $\langle x_\mathrm{HI}\rangle_m = 0.5$ at $z=7$. The leftmost column shows one grid-size thick slices of neutral hydrogen ($x_\mathrm{HI}\cdot (1+\delta_m)$), with halos within those slices scattered in pale green. These hybrid boxes were generated with grid size $\Delta x_\mathrm{grid}=0.625 h^{-1}$cMpc ($N_\mathrm{grid}=128^3$) using a tidal tensor-based matching criterion. The rightmost column shows the same for a reference simulation smoothed at the same resolution. Certain regions of interest are zoomed in and shown in the middle two columns.}
        \label{fig:HI_map_evolution}
    \end{figure*}

    \begin{figure*}
        \includegraphics[width=\textwidth]{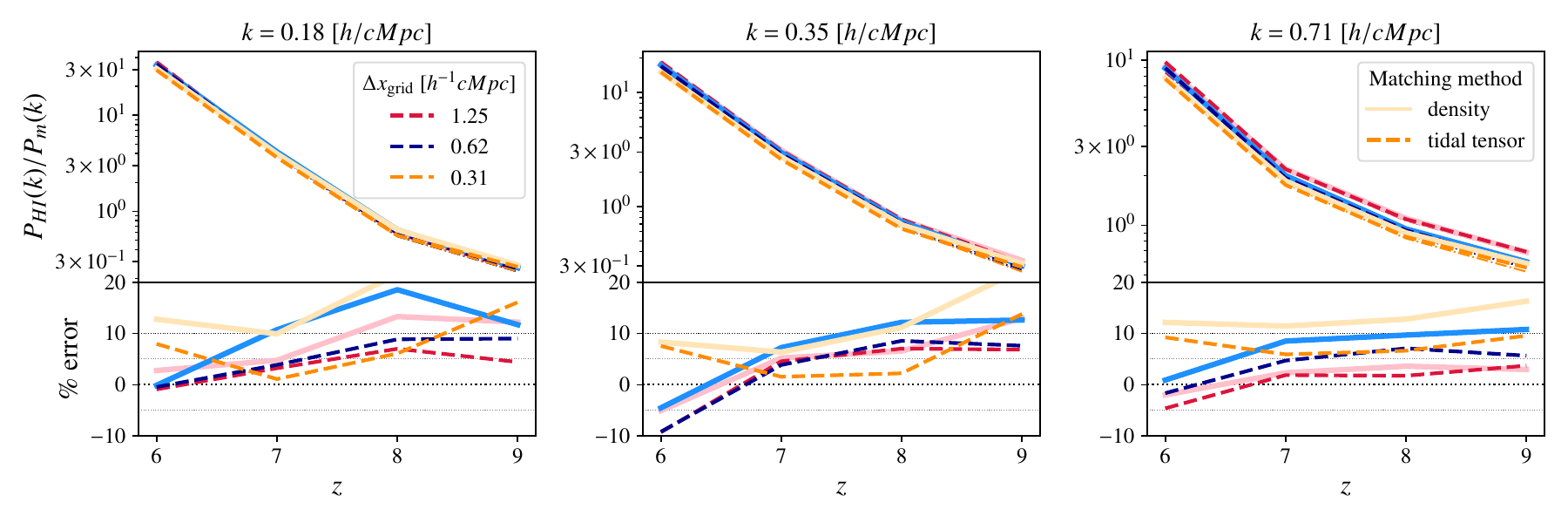}
        \caption{Power spectrum $P_\mathrm{HI}(k)$ of the HI density field scaled by the density power spectrum $P_m(k)$as a function of redshift for three wavenumbers $k$. Results are shown for three values of $\Delta x_\mathrm{grid}$ and two matching methods. The ionization efficiency $\zeta$ was chosen such that the reference box (RB) had a mass-averaged neutral fraction $\langle x_\mathrm{HI}\rangle_m = 0.5$ at $z=7$. \textit{Thin dash-dot lines} represent the HI power spectrum for the RB. The lower panel shows the \% error in each scaled power spectrum with respect to the RB.}
        \label{fig:HI_pspec_evolution}
    \end{figure*}

    Figure \ref{fig:hb_Ngrid_512_z_7.0_HI} shows the HI power spectra $P_\mathrm{HI}(k)$, scaled by $P_m(k)$, at $z=7$, obtained using our method for three values of mass-averaged neutral fraction. As in the case of power spectra of the haloes, we find that the HB built using the closeness of tidal tensor eigenvalues as the matching criteria match better with the RB. So let us focus on the results obtained using the match in tidal tensor eigenvalues (solid lines). For values of $\langle x_\mathrm{HI}\rangle_m < 0.75$, the agreement between the HB and RB is within $\sim 10\%$ for $k \lesssim 2.5 h /$~cMpc, irrespective of the grid size used. For $\langle x_\mathrm{HI}\rangle_m \gtrsim 0.75$, the disagreement exceeds $10\%$ for the smallest grid size $\Delta x_\mathrm{grid} = 0.3125 \hcMpc$. The disagreement seems to be noisy and could arise due to fewer ionized bubbles at these stages. This shows that there is some scope for improving the method for early stages of reionization and for smaller grid sizes. However, for the larger two grid sizes, the match is within $10\%$. We also find that for the largest grid size $\Delta x_\mathrm{grid} = 1.25 \hcMpc$ used in this work, the match is within $5 \%$ if we limit to relatively later stages of reionization $\langle x_\mathrm{HI}\rangle_m \lesssim 0.75$.
    In summary, we find that it can provide HI power spectra accurate to within $10 \%$ for $k \lesssim 2.5 h /$cMpc for grid size $\Delta x_\mathrm{grid} \geq 0.625 \hcMpc$.

    We next check the performance of our method at other redshifts. We choose five redshifts $z = 9, 8, 7, 6, 5$. The efficiency $\zeta$ is fixed at a value that leads to $50\%$ ionization at $z=7$ in the RB. This leads to a reionization history with $\langle x_\mathrm{HI}\rangle_m = 0.86, 0.73, 0.5, 0.14, 0.0$ at $z = 9, 8, 7, 6, 5$, respectively. The evolution of the HI density and halo field is shown in \fig{fig:HI_map_evolution} for $\Delta x_\mathrm{grid} = 0.625 \hcMpc$ for both HB and RB. As is clear from the comparison, the HB captures the large-scale distribution quite faithfully. There are some differences at smaller scales, which is expected as the halo populations in HB and RB are not identical at those scales.

    Finally, we show the evolution of $P_\mathrm{HI}(k) / P_m(k)$ with redshift for three values of $k$ in \fig{fig:HI_pspec_evolution}. We again find that their values for the HB and RB agree to within $10 \%$ for all the redshifts for $\Delta x_\mathrm{grid} \geq 0.625 \hcMpc$ when we use the closeness of tidal-tensor as the matching criterion for grid cells in LB and SB.

    \subsection{A different realization of SB}
    \label{sec:SB_10}
    The HB is designed to have an identical large-scale density field as the LB, which we in turn seed-matched with the RB for a faithful comparison devoid of cosmic variance effects. However, there is no such restriction on SB; it can be any realization of the dark matter density field so long that it has the correct box size and resolution. To check the robustness of our results against different realizations of SB, we generate another batch of HB with a different SB and plot its HI power spectra at $z=7$ in \fig{fig:HI_bias_SB_10}. The HI power spectra corresponding to the ``default'' SB are plotted in the same figure. While there are minor differences, overall we find that different random realizations of SB do not bias the results. In particular, the agreement between the RB and HB remains very similar for the two realizations of the SB. We have only shown the spectra of boxes built using the tidal-tensor matching method for clarity, however, the conclusion is the same even for hybrid boxes built using the density matching method.

    \subsection{Hybrid boxes with different sizes}
    \label{sec:L_40_160}

    \begin{figure*}[p]

        \includegraphics[width=\textwidth]{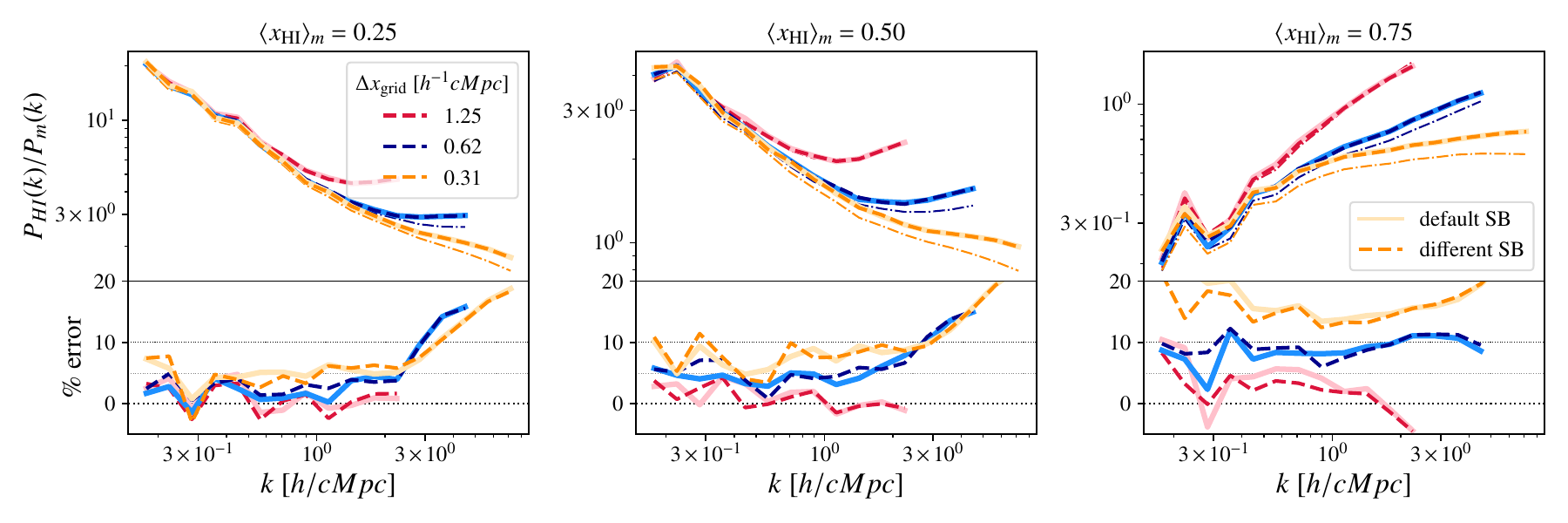}
        \caption{Power spectrum $P_\mathrm{HI}(k)$ of the HI density field scaled by the density power spectrum $P_m(k)$ at $z=7$ for two different hybrid boxes. The hybrid boxes were generated with the tidal tensor matching method using two different realizations of SB as indicated in the legend.}
        \label{fig:HI_bias_SB_10}

        \includegraphics[width=\textwidth]{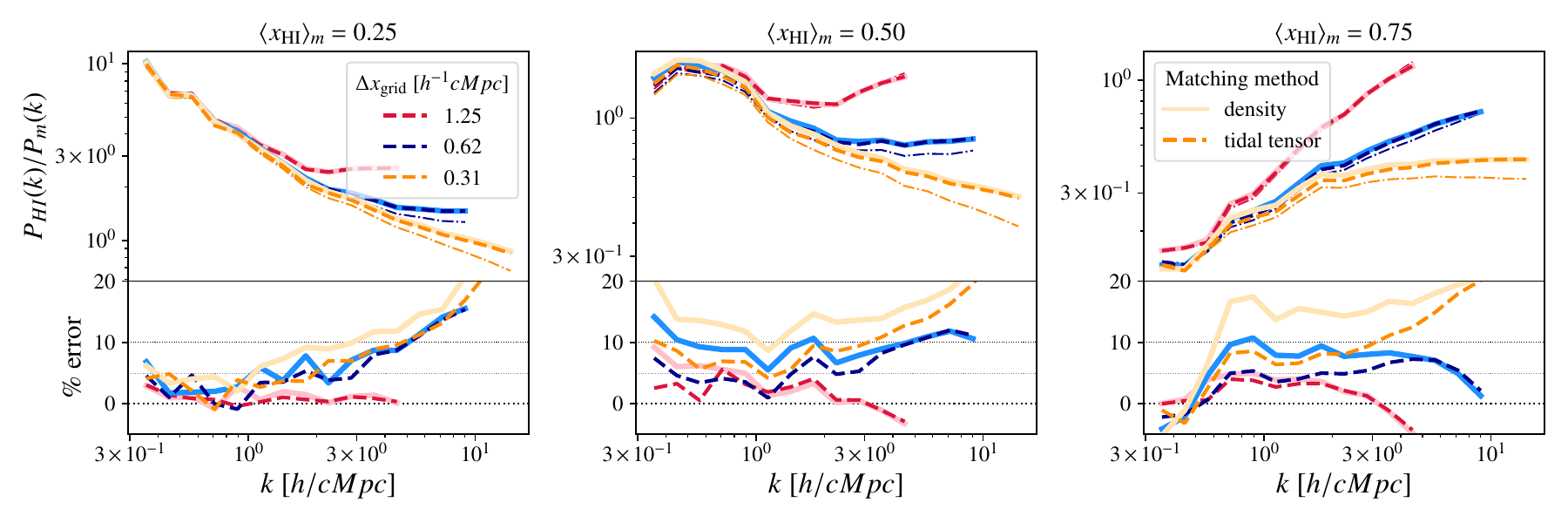}
        \caption{Same as \fig{fig:hb_Ngrid_512_z_7.0_HI} but for a HB of size $40 \hcMpc$ (half of the default box).}
        \label{fig:HI_bias_L_40}

        \includegraphics[width=\textwidth]{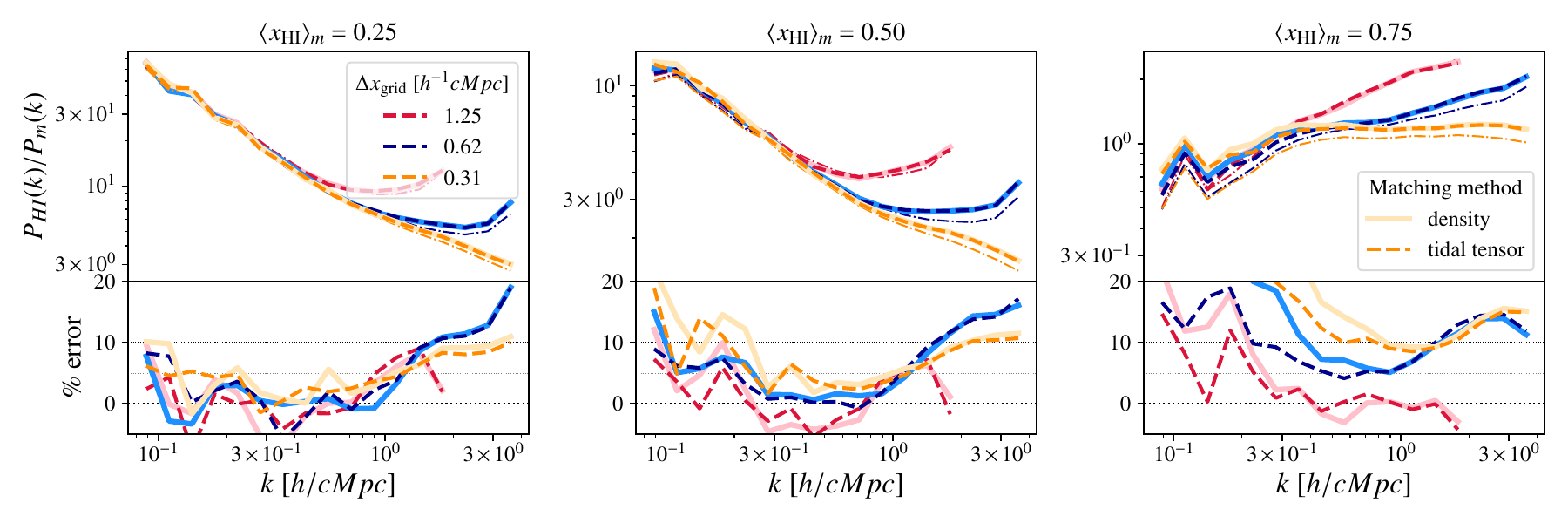}
        \caption{Same as \fig{fig:hb_Ngrid_512_z_7.0_HI} but for a HB of size $160 \hcMpc$ (twice of the default box).}
        \label{fig:HI_bias_L_160}

    \end{figure*}

    \begin{figure*}
        \includegraphics[width=\textwidth]{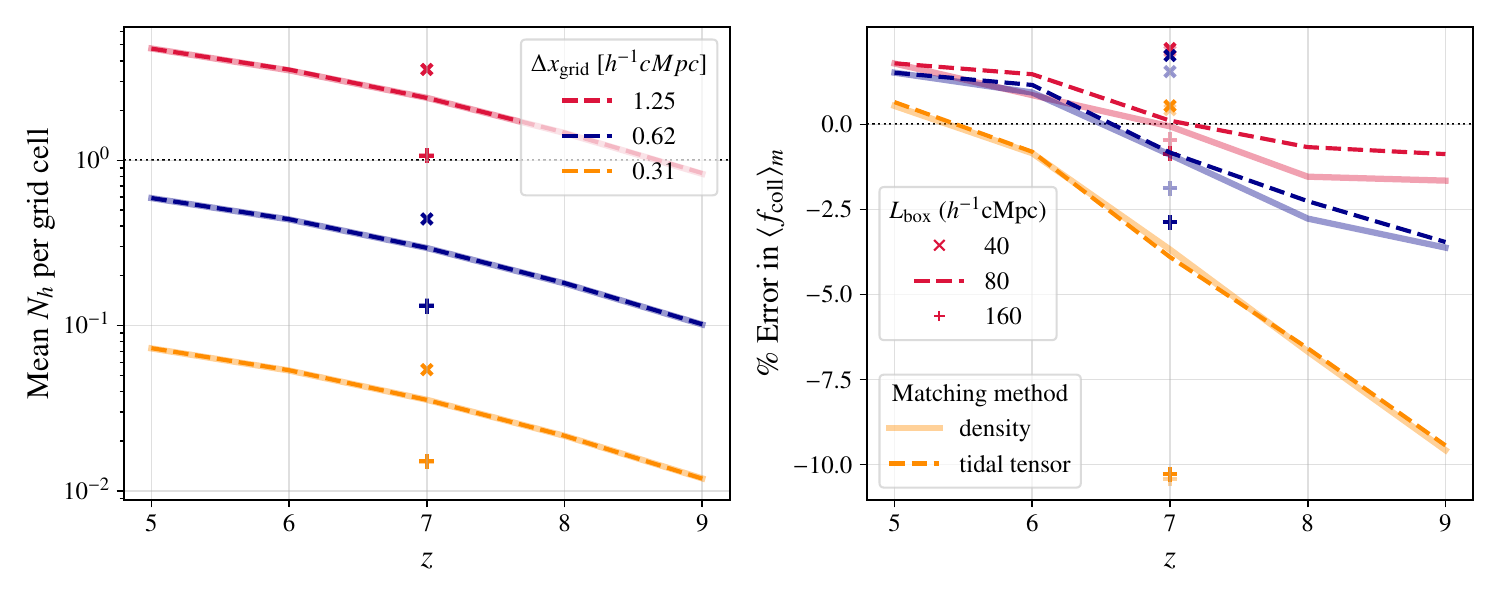}
        \caption{\textit{Left}: average number of haloes $N_h$ per grid cell in the different hybrid boxes used in this work. \textit{Right}: \% error in the mass-averaged collapsed fraction $\langle f_\mathrm{coll} \rangle_m$ in the HB with respect to RB. The results for the 40 and 160 $\hcMpc$ boxes are shown only at $z=7$. Note that there is little difference in the number of haloes between density and tidal tensor matched hybrid boxes, so the lines and scatterpoints overlap their fainter counterparts almost exactly in the left-hand panel.}
        \label{fig:Mean_Nhalo_err_fcoll}
    \end{figure*}

    In this section, we test our hybrid method at different scales using different box size. To this end, we construct the following two HBs:
    \begin{itemize}
        \item We build a smaller hybrid box of size $40 \hcMpc$ using LB and SB of sizes $20 \hcMpc$ and $40\hcMpc$ respectively each consisting of $512^3$ particles, and compare with a RB of size $40\hcMpc$ having $1024^3$ particles. The minimum halo mass $M_{h, \mathrm{min}} = 1.02 h^{-1} \Msun$ probed in this case is smaller than the default HB. The HI power spectra at $z = 7$ for this box is shown in \fig{fig:HI_bias_L_40}.

        \item We also build a larger hybrid box of size $160 \hcMpc$ using LB and SB of sizes $160 \hcMpc$ and $80\hcMpc$ respectively, each containing $512^3$ particles and compare with an appropriate RB. The minimum halo mass is $M_{h, \mathrm{min}} = 6.52 h^{-1} \Msun$ in this case. The HI power spectra at $z = 7$ are plotted in \fig{fig:HI_bias_L_160}.
    \end{itemize}

    Looking at Figures~\ref{fig:HI_bias_SB_10}, \ref{fig:HI_bias_L_40}, and \ref{fig:HI_bias_L_160} more closely, we notice two points: (i) At the same grid size, the mismatch between HB and RB increases as box size increases. (ii) The mismatch in a larger HB for a particular grid size is smaller than that in a smaller HB for the next smallest grid size. As an example, note that the HI power spectra for the $160 \hcMpc$ HB with grid size $0.62 \hcMpc$ are less accurate than those for the $80 \hcMpc$ box with grid size $0.62 \hcMpc$, but are more accurate than when the latter's grid size is $0.31 \hcMpc$.

    The above results can be explained by studying the fluctuations in the number of haloes per grid cell. The left panel in \fig{fig:Mean_Nhalo_err_fcoll} shows the mean number of haloes per grid cell in the hybrid boxes for different grid sizes and redshifts. Larger grid sizes lead to a larger number of haloes per cell for the same box size. The right panel of the same figure shows the errors in the mass averaged collapsed fraction $\langle f_\mathrm{coll} \rangle_m = \langle f_\mathrm{coll} \cdot (1+\delta_m) \rangle$.\footnote{The difference in $\langle f_\mathrm{coll} \rangle_m$ between HB and RB is the reason why we need to ``tune'' our ionization efficiency $\zeta$ to match the neutral fraction between the two, though the discrepancy is $\lesssim 5\%$ for $\Delta x_\mathrm{grid} \gtrsim 0.625 \hcMpc$. Note that $\langle x_\mathrm{HI} \rangle_m = 1 - \zeta \times \langle f_\mathrm{coll} \rangle_m$.} Comparing the two panels, it is clear that the mean halo numbers are correlated with the errors in $\langle f_\mathrm{coll} \rangle_m$. Since the HI fields are computed from the $f_\mathrm{coll}$ field, we would expect any errors in the latter to propagate to the former. This indeed seems to be the case as the errors in HI power spectra seem to get worse in the same sequence that one would encounter by following the mean number of haloes per cell \textit{downwards} at $z=7$, Thus, to minimize errors in HI power spectra, we recommend that the grid size be chosen such that there is a sufficient number of haloes per grid cell in the hybrid box, typically $\gtrsim 1$.

    \subsection{Computational Advantage}
    \label{sec:computational_advantage}

    The most significant advantage of our hybrid method is that we can obtain high-dynamic range information from low-dynamic range simulation boxes having much fewer particles, therefore requiring much less memory to run. The peak memory requirement for building the hybrid box from parent boxes is designed to never exceed that of loading one of those parent boxes in memory. Therefore, we can build the entire hybrid box using a computer that can only simulate
    \begin{equation}
        \frac{N_\mathrm{part}}{N_\mathrm{part, large}} = \frac{N_\mathrm{part}}{N_\mathrm{part, small}}
    \end{equation}
    ($= 8$ in our case) times fewer particles.

    Currently, our hybrid method requires parent simulations that are run with 8 times fewer particles than what a full simulation would require, thus reducing the memory requirement of \texttt{GADGET-2} simulations by 87\%. One could get away with an even smaller amount of memory at the cost of increasing the errors in the HI power spectra.

    Since the LB and SB both contain fewer particles than RB, the computational time taken for constructing the HB is significantly smaller. As mentioned in Table~\ref{tab:simulations}, we can generate the LB and SB in about $7$ times less time compared to the RB. Compared to the amount of time it takes to run the parent simulations, our hybrid method is nearly instantaneous, taking only a few minutes to run on a single CPU core.

    \section{Summary}
    \label{sec:summary}

    Numerical studies of reionization can be quite expensive due to the high dynamic range demanded by such simulations \cite{2022LRCA....8....3G,2022GReGr..54..102C}. It is therefore challenging to explore the space of unknown parameters when fitting theoretical models to data. While this challenge has been addressed to a large extent via computationally efficient semi-numerical simulations, they still have the shortcoming that small haloes, potentially hosting the first stars, cannot be resolved directly \cite{2007ApJ...669..663M,2011MNRAS.411..955M,2010MNRAS.406.2421S,2014MNRAS.443.2843M,2016MNRAS.457.1550H,2017MNRAS.464.2992M,2018MNRAS.477.1549H,2018MNRAS.481.3821C}. This problem is usually addressed using some sub-grid prescriptions to populate the simulation volume with unresolved haloes, e.g., those based on an analytical conditional halo mass function \cite{2009MNRAS.394..960C,2014MNRAS.440.1662S, 2008MNRAS.384.1069B,2007ApJ...654...12Z, 2011MNRAS.411..955M,2011A&A...527A..93S,2015MNRAS.447.1806G, 2022A&A...667A.118D}.

    In this work, we take an alternate hybrid approach to populate a low-resolution simulation volume with sub-grid haloes by combining information from two simulation boxes having relatively narrow dynamic ranges. Our hybrid scheme uses the density field from a coarse resolution simulation, having a size that can access the large scales required in the problem. The low-mass haloes cannot be resolved in this simulation, so another high-resolution simulation of smaller volume is run to resolve these haloes. We then match grid cells from the large box and the small box based on their physical properties (e.g., density and/or tidal environment) and populate haloes from the small box to the large one. Our method can provide an effective high dynamic range simulation by combining two boxes having relatively narrow dynamic ranges. An advantage of our method is that it does not require assuming any functional form for the underlying halo mass function. Instead, it is derived self-consistently from the simulations used, minimizing any uncertainty arising from the choice of the functional form \cite{2024arXiv240314061G}.

    While finding the matching cells between the large and small boxes, we explore two different criteria. The first is to find a matching cell that is closest in the density contrast $\delta_m$, and the second method uses the match between the eigenvalues of the tidal tensor. Our analyses show that the second method provides a better match between the hybrid and reference box, thus highlighting the importance of the tidal environment in determining the large-scale distribution of haloes.

    We find that, compared to a reference simulation done using the cosmological simulation \texttt{GADGET-2} and semi-numerical code \texttt{SCRIPT}, our hybrid method accurately reproduces the HI maps as well as summary statistics relevant to the studies of reionization. The hybrid halo mass function, large-scale halo auto-power spectra and halo-matter cross-power spectra, and the HI power spectra are all reproduced within $\sim 10 \%$ at scales $k \lesssim 2.5 h /$cMpc. Our method gives accurate results at a range of redshifts relevant to the studies of reionization, enabling detailed studies of the temporal evolution of the 21~cm signal. We also find that our method performs best when the number of haloes per grid cell in the hybrid box $\gtrsim 1$. This suggests improved accuracy with larger grid cell sizes which, however, comes at the expense of losing sensitivity to smaller scales (i.e., larger $k$-modes). Therefore, there exists an optimal grid cell size for any given simulation setup, balancing accuracy and scale sensitivity.

    Our method is at least 7 times faster than a full \texttt{GADGET-2} simulation having the equivalent dynamic range. More importantly, the hybrid box can be built with only $13\%$ of the memory required for a full simulation. In principle, our method can be even less expensive if one is only interested in the really large scales, and can afford some loss of accuracy at smaller scales. In such a case, boxes with 4 times narrower dynamic range (as opposed to 2 times narrower boxes that we have analyzed in this paper) can be explored. The caveat to consider in such a case is that the small box may not sample the extreme densities that are present in the large box, thus finding appropriately matched grid cells could be difficult. In addition, the largest halo mass reliably sampled in the small box might be smaller than the smallest halo resolved in the large box, thus leading to a gap in the hybrid halo mass function. Further analyses are required to quantify these effects.

    Encouraged by our success in accurately reproducing the two-point statistics of dark matter, halo, and HI density fields, we wish to extend our approach to incorporate the effects of redshift-space distortions on the power spectra. This would require building a hybrid velocity field that is well correlated with the density, halo, and HI fields, and we leave it for future work. We are also interested in measuring how well our hybrid method reproduces higher-order statistics such as three-point correlations of these fields.

    The efficiency of our method enables the creation of reasonably accurate halo catalogues with relatively fewer computational resources, which can be used to generate large-scale HI maps. This method can be particularly useful for varying cosmological parameters with semi-numerical reionization simulations, a traditionally challenging task \cite{kern2017emulating}. With recent advancements in techniques like simulation-based inference for constraining reionization models \cite{2022ApJ...926..151Z, 2022ApJ...933..236Z, 2023MNRAS.524.4239P, 2023MNRAS.525.6097S}, it is worth exploring how upcoming reionization observations can be leveraged to probe cosmological parameters effectively using our hybridization method.

    \section*{Acknowledgments}
    The authors are grateful to the reviewer for helpful comments that led to a better understanding of the caveats in our method, as well as improved the readability of this work. TRC acknowledges the insightful discussions with Aditya Chowdhury and Girish Kulkarni during the initial stages of this work.

    \section*{Data Availability}
    The data generated during this work will be made available upon reasonable request to the corresponding author.

    \bibliographystyle{JHEP}
    \bibliography{main}

    \appendix

    \label{lastpage}
\end{document}